\documentclass[11pt,draft,twocolumn]{report}

\usepackage{amsmath,amssymb}

\makeatletter
\newenvironment{harvard}{\list{}{\topsep=0\p@\parsep=0\p@
\partopsep=0\p@\itemsep=0\p@\labelsep=0\p@\itemindent=-18\p@
\labelwidth=0\p@\leftmargin=18\p@ }\normalsize\rm
\def\newblock{\ }
\sloppy\clubpenalty4000\widowpenalty4000
\sfcode`\.=1000\relax}{\endlist}
\def\refs{\begin{harvard}}
\def\endrefs{\end{harvard}}
\makeatother
\newcommand\ts{\mkern 2mu}
\newcommand\rmd{\mathrm{d}}

\DeclareMathOperator\End{{\mathrm {End}}}
\DeclareMathOperator\GL{{\mathrm {GL}}}
\DeclareMathOperator\half{\tfrac 12}
\begin{document}
\setcounter{page}{189}
\noindent{\sffamily\bfseries\large  EINSTEIN--CARTAN THEORY}
\bigskip

\noindent{\sffamily\bfseries Andrzej Trautman}, {\sffamily
Warsaw University, Warsaw, Poland}
\bigskip

\noindent{\small Published with typographical changes in:\newline
 Encyclopedia of Mathematical Physics,
edited by \newline J.-P. Fran\c{c}oise, G.L. Naber and Tsou S.T.\newline  Oxford:
Elsevier, 2006, vol. 2,   pages 189--195.}
\bigskip

\noindent{\sffamily\bfseries\large Introduction}
\medskip

\noindent{\sffamily\bfseries Notation}
\smallskip

Standard notation and terminology of differential geometry  and general relativity
are used in this article.
All considerations are local so that the
four-dimensional {\it space-time\/} \(M\) is assumed to be a smooth
manifold diffeomorphic to \(\mathbb{R}^4\). It is endowed with
a {\it metric tensor\/} \(g\) of signature \((1,3)\)   and
a {\it linear connection\/} defining the {\it covariant differentiation\/} of
tensor fields. Greek indices range from 0 to  3 and refer to  space-time.
Given  a field of frames \((e_\mu)\) on \(M\), and
the dual field of coframes \((\theta^\mu)\),
one can write the metric tensor as \(g=g_{\mu\nu}\theta^\mu\theta^\nu\),
where \(g_{\mu\nu}=g(e_\mu,e_\nu)\) and Einstein's summation convention is
assumed to hold. Tensor indices are lowered with \(g_{\mu\nu}\)
and raised with its inverse \(g^{\mu\nu}\).
General-relativistic units are used so that
 both Newton's constant of gravitation
and the speed of light  are 1. This implies   \( \hbar = \ell^2\),
where \(\ell\approx 10^{-33}\) cm is the Planck length. Both
mass and energy are measured in centimeters.
 \smallskip

\noindent{\sffamily\bfseries Historical remarks}
\smallskip

The Einstein--Cartan  Theory (ECT) of gravity is a modification of
General Relativity Theory (GRT),  allowing space-time
to have torsion, in addition to curvature, and relating torsion to
the density of intrinsic angular momentum.
This modification was put forward in 1922 by \'Elie Cartan, before the
discovery of  spin. Cartan was influenced by the work of the
Cosserat brothers (1909), who considered besides an (asymmetric)
force stress tensor also a moments stress tensor in a suitably
generalized continuous medium.
 Work done in the 1950s by physicists (Kondo, Bilby, Kr\"oner
and other authors) established the role played by torsion in the
continuum theory of crystal dislocations.
A recent review (Ruggiero and Tartaglia, 2003) describes the
links between ECT and the classical theory of defects in an elastic medium.

Cartan assumed the linear connection to be metric and derived, from a
variational principle, a set of gravitational field equations.
He required, without justification, that the covariant
divergence of the energy-momentum tensor be zero; this led to an
algebraic constraint equation, bilinear in curvature and torsion, severely
restricting the geometry. This misguided observation has probably
discouraged Cartan from pursuing his theory. It is now known that
conservation laws in relativistic  theories of gravitation follow
from the Bianchi identities and, in the presence of torsion, the
divergence of the energy-momentum tensor need not vanish.
Torsion is implicit in the 1928 Einstein  theory of gravitation with
teleparallelism.
 For a long time,  Cartan's modified theory of gravity, presented in his rather
abstruse notation, unfamiliar to physicists,   did not attract any attention.
 In the late 1950s, the theory of gravitation with spin and torsion
 was independently rediscovered by Sciama and Kibble. The role of Cartan
 was recognized soon afterwards and ECT became the
 subject of much research; see Hehl et al. (1976) for a review
 and an extensive bibliography. In the 1970s it was recognized that ECT
 can be incorporated within supergravity.
 In fact, simple supergravity is  equivalent to ECT with
 a massless, anticommuting
  Rarita--Schwinger field as the source. Choquet-Bruhat considered
 a generalization of ECT to higher dimensions  and showed that
the Cauchy problem for the coupled system of Einstein--Cartan and Dirac
equations is well posed.
 Penrose (1982) has shown  that
 torsion appears in a natural way
 when spinors are allowed to be rescaled by a {\it complex\/}
 conformal factor.
 ECT has been generalized by allowing non-metric linear connections and
 additional currents, associated with dilation and shear, as sources
of such a ``metric-affine theory of gravity'' (Hehl et al., 1995).

\smallskip

\noindent{\sffamily\bfseries Physical motivation}
\smallskip

Recall that, in Special Relativity Theory (SRT), the underlying
Minkowski space-time admits, as its group of automorphisms,
the full Poincar\'e group, consisting of translations {\it and\/}
 Lorentz transformations. It follows from the first Noether theorem
 that classical, special-relativistic field equations, derived from
 a variational principle, give rise to  conservation
 laws of energy-momentum {\it and\/} angular momentum. Using
 Cartesian coordinates \((x^\mu)\), abbreviating
 \(\partial\varphi/\partial x^\rho\) to \(\varphi_{,\rho
 }\) and denoting by
\(t^{\mu\nu}\) and \(s^{\mu\nu\rho}=-s^{\nu\mu\rho}\)
the tensors of energy-momentum and of intrinsic
  angular momentum (spin), respectively, one can write
  the conservation laws in the form
 \begin{equation}\label{e:tSRT}
    {t^{\mu\nu}}_{,\nu}=0
\end{equation}
and
\begin{equation}\label{e:sSRT}
    (x^\mu t^{\nu\rho}-x^\nu t^{\mu\rho} +{s^{\mu\nu\rho}})_{,\rho}=0.
\end{equation}
In the presence of spin, the tensor \(t^{\mu\nu}\)
need not be symmetric,
\begin{equation*}
    t^{\mu\nu}-t^{\nu\mu}={s^{\mu\nu\rho}}_{,\rho}.
\end{equation*}
Belinfante and Rosenfeld  have shown that
 the tensor
\begin{equation*}
    T^{\mu\nu}=t^{\mu\nu}+
    \half(s^{\nu\mu\rho}+s^{\nu\rho\mu}+s^{\mu\rho\nu})_{,\rho}
\end{equation*}
is symmetric and its divergence vanishes.

In quantum theory,
the irreducible, unitary representations
of the Poincar\'e group correspond to elementary  systems such as
stable particles;
these representations are labeled by the mass {\it and\/} spin.

In Einstein's GRT, the space-time \(M\) is curved;  the Lorentz group --- but
not the Poincar\'e group --- appears as the structure group acting on
orthonormal frames in the tangent spaces of \(M\). The energy-momentum
tensor \(T\) appearing on the right side
of the Einstein equation is necessarily symmetric. In GRT there is no
room for translations and  the tensors \(t\) and \(s\).

By introducing torsion and relating it to \(s\), Cartan restored the
role of the Poincar\'e group in relativistic gravity: this group acts on
the affine frames in the tangent spaces of \(M\). Curvature and torsion
are the surface densities of Lorentz transformations and translations, respectively.
In a space with torsion, the Ricci tensor need not be symmetric so that
an asymmetric energy-momentum tensor
can appear on the right side of the Einstein equation.
\medskip

\noindent{\sffamily\bfseries\large  Geometric preliminaries}
\medskip

\noindent{\sffamily\bfseries Tensor-valued differential
forms}
\smallskip

It is convenient to follow Cartan in describing geometric objects
as tensor-valued differential forms. To define them,
consider a homomorphism
\(\sigma:\GL_4(\mathbb{R})\to\GL_N(\mathbb{R})\) and an
element \(A=(A^\mu_\nu)\) of \(\End \mathbb{R}^4\), the
Lie algebra of \(\GL_4(\mathbb{R})\).
The derived representation of Lie algebras is given by
\(
\tfrac{\rmd}{\rmd t}\sigma(\exp At){\vert_{t=0}}=\sigma_\mu^\nu A^\mu_\nu.
\)
 If \((e_a)\) is a frame in \(\mathbb{R}^N\),
  then
  \(\sigma_\mu^\nu(e_a)=\sigma_{a\mu}^{b\nu}e_b\), where
\(a,b=1,\dots,N\).

A map \(a=({a^\mu}_\nu):M\to\GL_4(\mathbb{R})\) transforms
 fields of frames so that
\begin{equation}\label{e:trfr}
    e'_\mu=e_\nu a^\nu_\mu\quad\mbox{and}\quad \theta^\nu=a^\nu_\mu\theta'{^\mu}.
\end{equation}
 A differential form \(\varphi\) on \(M\),
with values in \(\mathbb{R}^N\), is said to be of type \(\sigma\) if,
under changes of frames, it transforms so that \(\varphi'=\sigma(a^{-1})\varphi\).
For example, \(\theta =(\theta^\mu)\) is a 1-form of type id.
If now \(A=(A^\mu_\nu):M\to\End \mathbb{R}^4\), then one puts  \(a(t)=
\exp tA:M\to\GL_4(\mathbb{R})\) and  defines the variations
induced by an infinitesimal change of frames,
\begin{equation}\label{e:defd}
\begin{split}\delta \theta &
=\frac{\rmd}{\rmd t}({a(t)}^{-1}\theta){\vert_{t=0}}
=-A\theta,\\
    \delta \varphi &
    =\frac{\rmd}{\rmd t}(\sigma({a(t)}^{-1})\varphi){\vert_{t=0}}
    =-\sigma_\mu^\nu A^\mu_\nu\varphi.
\end{split}
\end{equation}
\smallskip

\noindent{\sffamily\bfseries Hodge duals}
\smallskip

Since \(M\) is diffeomorphic to \(\mathbb{R}^4\), one can  choose
an orientation on \(M\) and restrict the frames to
agree with that orientation so that only transformations with values in
\(\GL^+_4(\mathbb{R})\) are allowed. The metric then defines
the Hodge dual of differential forms. Put \(\theta_\mu=g_{\mu\nu}\theta^\nu\).
The forms \(\eta\), \(\eta_\mu\), \(\eta_{\mu\nu}\),
\(\eta_{\mu\nu\rho}\) and \(\eta_{\mu\nu\rho\sigma}\)
are defined to be the duals of \(1\), \(\theta_\mu\), \(\theta_\mu\wedge\theta_\nu\),
\(\theta_\mu\wedge\theta_\nu\wedge\theta_\rho\) and
\(\theta_\mu\wedge\theta_\nu\wedge\theta_\rho\wedge\theta_\sigma\), respectively.
The 4-form \(\eta\) is the volume element; for a holonomic coframe
\(\theta^\mu=\rmd x^\mu\) it is given by \(\sqrt{-\det(g_{\mu\nu})}\,\rmd x^0
\wedge\rmd x^1\wedge\rmd x^2\wedge\rmd x^3\).
In SRT, in Cartesian coordinates, one can  define the tensor-valued 3-forms
\begin{equation}\label{e:defts}
    t^\mu=t^{\mu\nu}\eta_\nu\quad\mbox{and}\quad s^{\mu\nu}=s^{\mu\nu\rho}\eta_\rho
\end{equation}
so that equations \eqref{e:tSRT} and \eqref{e:sSRT} become
\begin{equation*}
    \rmd t^\mu=0\quad\mbox{and}\quad \rmd j^{\mu\nu}=0,
\end{equation*}
where
\begin{equation}\label{e:defj}
    j^{\mu\nu}=x^\mu t^\nu-x^\nu t^\mu+s^{\mu\nu}.
\end{equation}
  For an isolated system, the 3-forms \(t^\mu\) and \(j^{\mu\nu}\),
integrated over the 3-space \(x^0=\)const.,
give the system's total energy-momentum vector and angular momentum
bivector, respectively.

\smallskip

\noindent{\sffamily\bfseries Linear connection, its curvature and torsion}
\smallskip

A linear connection on \(M\) is represented, with respect to
the field of frames, by the field of 1-forms
\begin{equation*}
    \omega^\mu_\nu=\varGamma^\mu_{\rho\nu}\theta^\rho
\end{equation*}
so that the covariant derivative of \(e_\nu\) in the direction of
\(e_\mu\) is \(\nabla_\mu e_\nu=\varGamma^\rho_{\mu\nu}e_\rho\).
Under a change of frames \eqref{e:trfr}, the connection forms transform
as follows:
\begin{equation*}
    a^\mu_\rho\omega'^\rho_\nu=\omega^\mu_\rho a^\rho_\nu+\rmd a^\mu_\nu.
\end{equation*}
If \(\varphi=\varphi^a e_a\) is a \(k\)-form of type \(\sigma\), then
its {\it covariant exterior derivative\/}
\begin{equation*}
    D\varphi^a=\rmd\varphi^a+\sigma^{a\mu}_{b\nu}\omega^\nu_\mu\wedge\varphi^b
\end{equation*}
is a \((k+1)\)-form of the same type.  For a \(0\)-form
one has \(D\varphi^a=\theta^\mu\nabla_\mu\varphi^a\).
The infinitesimal change of \(\omega\), defined similarly as in \eqref{e:defd},
is
\(\delta\omega^\mu_\nu=DA^\mu_\nu\).
The 2-form of curvature \(\varOmega=({\varOmega^\mu}_\nu)\), where
\begin{equation*}
    \varOmega^\mu_{\ \nu}=\rmd\omega^\mu_\nu+\omega^\mu_\rho\wedge
\omega^\rho_\nu,
\end{equation*}
is  of type ad: it transforms with the adjoint representation
of \(\GL_4(\mathbb{R})\) in \(\End\mathbb{R}^4\).
The 2-form of torsion \(\varTheta=(\varTheta^\mu)\), where
\begin{equation*}
    \varTheta^\mu=\rmd\theta^\mu+\omega^\mu_\nu\wedge\theta^\nu,
\end{equation*}
is of type id. These forms
  satisfy the {\it Bianchi identities}
\begin{equation*}
  D\varOmega^\mu_{\ \nu}=0\quad\mbox{and}\quad D\varTheta^\mu=
  \varOmega^\mu_{\ \nu}\wedge\theta^\nu.
\end{equation*}
  For a differential form \(\varphi\) of type \(\sigma\) there
  holds the identity
  \begin{equation}\label{e:D2}
    D^2\varphi^a=\sigma^{a\nu}_{b\mu}\varOmega^\mu_{\ \nu}\wedge\varphi^b.
\end{equation}
The  tensors of curvature and torsion are given by
\begin{equation*}
\varOmega^\mu_{\ \nu}=\half R^\mu_{\ \nu\rho\sigma}\theta^\rho
\wedge\theta^\sigma
\end{equation*}
and
\begin{equation*}
    \varTheta^\mu=\half Q^\mu_{\ \rho\sigma}\theta^\rho\wedge\theta^\sigma,
\end{equation*}
respectively.
With respect to a holonomic frame, \(\rmd \theta^\mu=0\),  one has
\begin{equation*}
    Q^\mu_{\ \rho\sigma}=\varGamma^\mu_{\rho\sigma}-\varGamma^\mu_{\sigma\rho}.
\end{equation*}
In SRT, the Cartesian coordinates define a radius-vector field
\(X^\mu=-x^\mu\), pointing towards the origin of the coordinate system.
The differential equation it satisfies
generalizes to a manifold with a linear connection:
\begin{equation}\label{e:defrv}
    DX^\mu+\theta^\mu=0.
\end{equation}
By virtue of \eqref{e:D2},
 the integrability condition of \eqref{e:defrv} is
 \begin{equation*}
   \varOmega^\mu_{\ \nu}X^{\nu}+\varTheta^{\mu}=0.
\end{equation*}
 Integration
of \eqref{e:defrv} along a curve defines the Cartan displacement of
\(X\); if this is done along a small closed circuit spanned by the
bivector \( \varDelta f \), then the radius vector changes by about
\begin{equation*}
\varDelta X^{\mu}=\tfrac{1}{2} (R^\mu_{\ \nu\rho\sigma}X^{\nu}
+Q^\mu_{\ \rho\sigma})\varDelta f^{\rho\sigma}.
\end{equation*}
This holonomy theorem --- rather imprecisely formulated here --- shows
that torsion bears to translations a relation similar to that of
curvature to linear homogeneous transformations.

In a space with torsion it matters whether one considers the
 potential of the electromagnetic field to be a scalar-valued 1-form
\(\varphi\) or a covector-valued 0-form \((\varphi_\mu)\).
The first choice leads to a field \(\rmd\varphi\) that is invariant
with respect to the gauge transformation \(\varphi\mapsto\varphi+\rmd\chi\).
The second gives
\(\half(\nabla_\mu\varphi_\nu-\nabla_\nu\varphi_\mu)\ts\theta^\mu\wedge
\theta^\nu=(D\varphi_\mu)\wedge\theta^\mu=\rmd\varphi-\varphi_\mu\varTheta^\mu\),
a gauge-dependent field.
\smallskip

\noindent{\sffamily\bfseries Metric-affine geometry}
\smallskip

A  {\it metric-affine\/} space \((M,g,\omega)\) is defined to have
a metric and a linear connection that need not dependent on each other.
The metric alone determines the torsion-free  Levi-Civita connection
 \(\mathring{\omega}\)
characterized by
\begin{equation*}
    \rmd\theta^\mu+\mathring{\omega}^\mu_\nu\wedge\theta^\nu=
    0\quad\mbox{and}\quad
    \mathring{D} g_{\mu\nu}=0.
    \end{equation*}
Its curvature is
\begin{equation*}
    {\mathring\varOmega}^\mu_{\ \nu}=\rmd\mathring{\omega}^\mu_\nu
    +\mathring{\omega}^\mu_\rho\wedge\mathring{\omega}^\rho_\nu.
\end{equation*}
The 1-form of type ad,
\begin{equation}\label{e:defka}
    \kappa^\mu_{\ \nu}=\omega^\mu_\nu-\mathring{\omega}^\mu_\nu,
\end{equation}
 determines the torsion of \(\omega\)
and the covariant derivative of \(g\),
\begin{equation*}
    \varTheta^\mu=\kappa^\mu_{\ \nu}\wedge\theta^\nu,\quad
    Dg_{\mu\nu}=-\kappa_{\mu\nu}-\kappa_{\nu\mu}.
\end{equation*}
The curvature of \(\omega\) can be written as
\begin{equation}\label{e:split}
    \varOmega^\mu_{\ \nu}=\mathring{\varOmega}^\mu_{\ \nu}+
    \mathring{D}\kappa^\mu_{\ \nu}+\kappa^\mu_{\ \rho}
    \wedge\kappa^\rho_{\ \nu}.
\end{equation}
 The {\it transposed\/} connection \(\tilde{\omega}\) is
defined by
\begin{equation*}
    \tilde{\omega}^\mu_\nu=\omega^\mu_\nu+Q^\mu_{\ \nu\rho}\theta^\rho
\end{equation*}
so that, with respect to a holonomic frame, one has
\(\tilde{\varGamma}^\mu_{\nu\rho}= \varGamma^\mu_{\rho\nu}\).
The torsion of \(\tilde{\omega}\) is opposed to that of \(\omega\).
\newpage

\noindent{\sffamily\bfseries Riemann--Cartan geometry}
\smallskip

A {\it Riemann--Cartan\/} space is a metric-affine space
with a connection that is metric,
\begin{equation}\label{e:metr}
    Dg_{\mu\nu}=0.
\end{equation}
The
metricity condition   implies \(\kappa_{\mu\nu}+\kappa_{\nu\mu}=0\)
and \(\varOmega_{\mu\nu}+\varOmega_{\nu\mu}=0\).
In a  Riemann--Cartan space the connection is determined
by its torsion \(Q\) and the metric tensor.
Let
\(Q_{\rho\mu\nu}=g_{\rho\sigma}Q^\sigma_{\ \mu\nu}\), then
\begin{equation}\label{e:relkQ}
\kappa_{\mu\nu}=\half(Q_{\mu\sigma\nu}
+Q_{\nu\mu\sigma}+Q_{\sigma\mu\nu})\theta^\sigma.
\end{equation}
The transposed connection of a Riemann--Cartan space is metric if, and only if, the
tensor  \(Q_{\rho\mu\nu}\)
is completely antisymmetric.
Let \(\tilde{\nabla}\) denote the covariant
 derivative with respect to \(\tilde{\omega}\).
 By definition, a {\it symmetry\/} of a Riemann--Cartan space
 is a diffeomorphism of \(M\) preserving
both \(g\) and \(\omega\). The one-parameter group
of local transformations of \(M\), generated by the vector field
\(v\), consists of symmetries of \((M,g,\omega)\) if, and only if,
\begin{equation}\label{e:Kill}
   \tilde{\nabla}^\mu v^\nu+\tilde{\nabla}^\nu v^\mu=0
\end{equation}
and
\begin{equation}\label{e:Lich}
    D\tilde{\nabla}_\nu v^\mu+R^\mu_{\ \nu\rho\sigma}v^\rho\theta^\sigma=0.
\end{equation}
In a Riemannian space, the connections \(\omega\) and \(\tilde{\omega}\)
coincide and \eqref{e:Lich} is a consequence of the Killing equation \eqref{e:Kill}.
The metricity condition implies
\begin{equation}\label{e:Deta}
    D\eta_{\mu\nu\rho}=\eta_{\mu\nu\rho\sigma}\varTheta^\sigma.
\end{equation}
\medskip

\noindent{\sffamily\bfseries\large The Einstein--Cartan theory of gravitation}
\medskip

\noindent{\sffamily\bfseries An identity resulting from local invariance}
\smallskip

Let \((M,g,\omega)\) be a metric-affine space-time.
Consider a Lagrangian \( L \) which is an invariant 4-form on \( M\), depends on
 \(g\), \(\theta\), \(\omega\), \(\varphi\), and the  first
 derivatives of \(\varphi=\varphi^a e_a\).
   The general variation of the Lagrangian is
 \begin{equation} \label{e:varL}
 \begin{split}
  \delta L&= L_{a}\wedge\delta\varphi^{a} +
 \tfrac
{1}{2} \tau^{\mu\nu}\delta g_{\mu\nu}
 +\delta\theta^{\mu}\wedge t_{\mu}\\
 &\qquad-\tfrac{1}{2}
\delta{\omega}^{\mu}_{\nu}
 \wedge s^{\nu}_{\ \mu} +\mbox{an exact form}
 \end{split}
 \end{equation}
 so that \(  L_{a}=0\)
 is the Euler--Lagrange equation for \( \varphi
 \).  If the changes of the
 functions \( \theta \), \( \omega \), \( g \) and \( \varphi  \) are
 induced by an infinitesimal change  of the frames \eqref{e:defd}
 then
 \( \delta L =0\) and \eqref{e:varL} gives  the identity
 \begin{equation*}
 g_{\mu\rho}\tau^{\rho\nu}-\theta^{\nu}\wedge t_{\mu} +\tfrac{1}{2}
 Ds^{\nu}_{\ \mu} -
 \sigma^{b\nu}_{a\mu}L_{a}\wedge\varphi^{b}=0.
 \end{equation*}
 It follows from the identity that the two sets
 of Euler--Lagrange equations obtained by varying \(
 L \) with respect to the triples \( (\varphi,\theta, \omega )\) and
 \( (\varphi, g,\omega  )\) are equivalent. In the sequel, the first
triple is chosen to derive the field equations.
\smallskip

\noindent{\sffamily\bfseries Projective transformations and the
metricity condition}
\smallskip

 Still under the assumption that \((M,g,\omega)\) is
  a metric-affine space-time, consider
  the 4-form
\begin{equation}\label{e:HEC}
8\pi K=\half g^{\nu\rho}\eta_{\mu\rho}\wedge\varOmega^{\mu}_{\ \nu}
\end{equation}
which is equal to  \(\eta R\), where
 \( R=g^{\mu\nu}R_{\mu\nu} \) is the Ricci scalar;
  the Ricci tensor \( R_{\mu\nu}=R^{\rho}_{\mu\rho\nu} \) is, in
general, asymmetric.  The form
 \eqref{e:HEC} is invariant with respect to {\it projective
 transformations\/} of the connection,
\begin{equation}
\omega^{\mu}_{\nu}\mapsto\omega^{\mu}_{\nu}+
\delta^{\mu}_{\nu}\lambda, \label{e:prt}
\end{equation}
where \(\lambda \) is an arbitrary 1-form. Projectively related
connections have the same (un\-para\-metrized) geodesics.
  If the
  total Lagrangian for gravitation interacting with the matter
  field \( \varphi \) is \(K+L\), then the field equations,
  obtained by varying it with respect to \( \varphi \), \( \theta
  \) and \(\omega \) are: \( L_{a}=0 \),
  \begin{equation}
\tfrac{1}{2}g^{\rho\sigma}\eta_{\mu\nu\rho}\wedge\varOmega^{\nu}_{\ \sigma}=
  -8\pi t_{\mu},
  \label{e:E1}
  \end{equation}
 and
 \begin{equation}
 D(g^{\mu\rho}\eta_{\rho\nu})=8\pi s^{\mu}_{\ \nu},
\label{e:C1}
 \end{equation}
 respectively. Put \(s_{\mu\nu}=g_{\mu\rho}s^\rho_{\ \nu}\). If
 \begin{equation}
 \label{e:sant}
  s_{\mu\nu}+s_{\nu\mu}=0,
  \end{equation}
then \( s^{\nu}_{\ \nu}=0 \) and \( L \) is also invariant with respect to
\eqref{e:prt}. One shows that, if \eqref{e:sant} holds, then,
among the projectively related connections satisfying
\eqref{e:C1}, there  is precisely one that is metric. To implement properly
the    metricity condition in the variational principle one can
use the Palatini approach with constraints (Kopczy\'nski, 1975).
Alternatively, following Hehl, one can use \eqref{e:defka} and \eqref{e:relkQ}
to eliminate \(\omega\) and obtain a lagrangian
 depending on \( \varphi, \theta  \) and
the tensor of torsion.

\smallskip

\noindent{\sffamily\bfseries The Sciama--Kibble field equations}
\smallskip

From now on the metricity condition \eqref{e:metr}
is assumed so that \eqref{e:sant} holds
and the Cartan field equation \eqref{e:C1} is
  \begin{equation}\label{e:C2}
  \eta_{\mu\nu\rho}\wedge\varTheta^{\rho}=8\pi s_{\mu\nu}.
  \end{equation}
 Introducing the asymmetric energy-momentum tensor \( t_{\mu\nu} \)
 and the spin density tensor \(  s_{\mu\nu\rho}=g_{\rho\sigma}
 s^\sigma_{\ \mu\nu}\) similarly as in \eqref{e:defts},
one can write the Einstein--Cartan equations \eqref{e:E1} and
\eqref{e:C2} in the  form
 given  by Sciama  and Kibble,
\begin{align}
R_{\mu\nu}-\tfrac{1}{2} g_{\mu\nu}R &=8\pi
t_{\mu\nu}
\label{e:Ein}\\
Q^{\rho}_{\ \mu\nu}+\delta^{\rho}_{\mu}Q^{\sigma}_{\ \nu\sigma}-
\delta^{\rho}_{\nu}Q^{\sigma}_{\ \mu\sigma} &= 8\pi
s^{\rho}_{\ \mu\nu}. \label{e:Car}
\end{align}
Equation  \eqref{e:Car} can be solved to give
\begin{equation}\label{e:Q=}
Q^{\rho}_{\ \mu\nu}=8\pi(s^{\rho}_{\ \mu\nu}
+\half\delta^\rho_\mu s^\sigma_{\ \nu\sigma}
+\half\delta^\rho_\nu s^\sigma_{\ \sigma\mu})
\end{equation}
Therefore, torsion vanishes in the absence
of spin and then \eqref{e:Ein} is the classical Einstein field
equation. In particular, there is no difference between the
Einstein and Einstein--Cartan theories in empty space. Since
practically all tests of relativistic gravity are based on
consideration of Einstein's equations in empty space, there is no
difference, in this respect, between the Einstein and the
Einstein--Cartan theories: the latter is as viable as the former.

In any case, the consideration of torsion amounts  to a slight
change of the energy-momentum tensor that can be also obtained by
the introduction of a new term in the Lagrangian. This observation
was made in 1950 by Weyl  in the context of the Dirac
equation.

In Einstein's theory one can also  satisfactorily describe
spinning matter without introducing torsion (Bailey and Israel, 1975).
\bigskip

\noindent{\sffamily\bfseries Consequences of the Bianchi identities: conservation laws}
\smallskip

Computing the covariant exterior derivatives of both  sides
of the Einstein--Cartan equations, using
\eqref{e:Deta} and
 the Bianchi identities
one obtains
\begin{equation}\label{e:CL1}
    8\pi Dt_\mu=\half\eta_{\mu\nu\rho\sigma}\varTheta^\nu\wedge\varOmega^{\rho\sigma}
\end{equation}
and
\begin{equation}\label{e:CL2}
8\pi Ds_{\mu\nu}=\eta_{\nu\sigma}\wedge\varOmega^{\sigma}_{\ \mu}-
\eta_{\mu\sigma}\wedge\varOmega^{\sigma}_{\ \nu}.
\end{equation}
Cartan required the right side of \eqref{e:CL1} to vanish.
If, instead, one uses the field equations
\eqref{e:E1} and  \eqref{e:C2} to evaluate the right sides of \eqref{e:CL1}
and \eqref{e:CL2}, one obtains,
\begin{equation}\label{e:Dt}
    Dt_\mu=Q^\rho_{\ \mu\nu}\theta^\nu\wedge t_\rho-
    \half R^\rho_{\ \sigma\mu\nu}\theta^\nu
    \wedge s^\sigma_{\ \rho}
\end{equation}
and
\begin{equation}\label{e:Ds}
    Ds_{\mu\nu}=\theta_\nu\wedge t_\mu-\theta_\mu\wedge t_\nu.
\end{equation}

Let \(v\) be a vector field generating a group of symmetries of the
Riemann--Cartan space \((M,g,\omega)\) so that equations \eqref{e:Kill}
and \eqref{e:Lich} hold.
Equations \eqref{e:Dt} and \eqref{e:Ds} then imply that the 3-form
\begin{equation*}
    j=v^\mu t_\mu +\half\tilde{\nabla}^\nu v^\mu s_{\mu\nu}
\end{equation*}
is closed, \(\rmd j=0\). In particular, in the limit of SRT,
in Cartesian coordinates \(x^\mu\), to a constant vector field \(v\)
there corresponds the projection onto \(v\) of the energy-momentum
density. If \(A^{\mu\nu}\) is a constant bivector,
 then \(v^\mu=A^\mu_{\ \nu} x^\nu\)
gives \(j=j^{\mu\nu}A_{\mu\nu}\), where \(j^{\mu\nu}\) is as in \eqref{e:defj}.
\smallskip

\noindent{\sffamily\bfseries Spinning fluid and
the generalized Mathisson--Papapetrou equation of motion}
\smallskip

As in classical general relativity, the right sides of the
Einstein--Cartan equations need not necessarily be derived from a
variational principle; they may be determined by phenomenological
considerations.
  For example, following Weyssenhoff,
 consider  a spinning fluid  characterized by
 \begin{equation*}
  t^{\mu\nu}=
 P^{\mu}u^{\nu} \quad\mbox{and}\quad
 s^{\mu\nu\rho}=S^{\mu\nu}u^{\rho},
\end{equation*}
where
\(S^{\mu\nu}+S^{\nu\mu}=0\)
 and \(u\) is the unit, timelike velocity field.
Let \(U=u^\mu\eta_\mu\) so that
\begin{equation*}
    t_\mu=P_\mu U\quad\mbox{and}\quad s_{\mu\nu}=S_{\mu\nu}U.
\end{equation*}
 Define the {\it particle derivative\/}
of a tensor field \(\varphi^a\)
 in the direction of  \( u \)  by
 \begin{equation*}
    \dot{\varphi}^a\eta=D({\varphi}^a U).
\end{equation*}
For a scalar field \(\varphi\), the equation \(\dot{\varphi}=0\)
is equivalent to the conservation law \(\rmd(\varphi U)=0\).
Define \(\rho=g_{\mu\nu}P^\mu u^\nu\), then \eqref{e:Ds} gives
an equation of motion of spin
\begin{equation*}
    \dot{S}_{\mu\nu}=u_\nu P_\mu-u_\mu P_\nu
\end{equation*}
so that
\begin{equation*}
P_{\mu}=\rho
u_{\mu}+\dot{S}_{\mu\nu}u^{\nu}.
\end{equation*}
From \eqref{e:Dt}  one obtains
the equation of translatory motion,
\begin{equation*}
\dot{P}_{\mu}=(Q^{\rho}_{\ \mu\nu}P_{\rho}-\tfrac{1}{2}
R^{\rho\sigma}_{\ \ \mu\nu} S_{\rho\sigma})u^{\nu},
\end{equation*}
which is a generalization to the Einstein--Cartan theory of the
Mathisson--Papapetrou equation for point particles with an
intrinsic angular momentum.
\smallskip

\noindent{\sffamily\bfseries From ECT to GRT: the effective energy-momentum tensor}
\smallskip

Inside spinning matter, one can use \eqref{e:relkQ} and \eqref{e:Q=}
to eliminate torsion  and replace the Sciama--Kibble
 system by a single Einstein equation with an effective
 energy-momentum tensor on the right side. Using the split \eqref{e:split}
 one can write \eqref{e:Ein} as
\begin{equation}\label{e:Eeff}
\mathring{R}_{\mu\nu}-\half g_{\mu\nu}\mathring{R}=8\pi T^{\rm eff}_{\mu\nu}.
\end{equation}
Here \(\mathring{R}_{\mu\nu}\) and \(\mathring{R}\) are, respectively,
 the Ricci tensor and scalar formed from \(g\).
The term in \eqref{e:split} that is quadratic in \(\kappa\) contributes
to \(T^{\rm eff}\) an expression quadratic in the components
of the tensor \(s_{\mu\nu\rho}\) so that,
neglecting indices, one can write symbolically
 \begin{equation} \label{e:eff}
   T^{\rm eff}=T+s^{2}.
\end{equation}
The symmetric tensor \(T\) is the sum of \(t\) and a term
coming from \(\mathring{D}\kappa^\mu_{\ \nu}\) in \eqref{e:split},
\begin{equation}\label{e:T}
    T^{\mu\nu}=t^{\mu\nu} +\half\mathring{\nabla}_\rho(
   s^{\nu\mu\rho}+s^{\nu\rho\mu}+s^{\mu\rho\nu}).
\end{equation}
It is remarkable that the  Belinfante--Rosenfeld symmetrization
of the canonical energy-momentum tensor appears as
a natural consequence of ECT. From the physical point
of view, the second term on the right side of \eqref{e:eff},
can be thought of as providing a spin-spin contact interaction,
reminiscent of the one appearing in the  Fermi theory of
weak interactions.

It is clear from \eqref{e:Eeff}, \eqref{e:eff} and \eqref{e:T}
 that whenever terms
 quadratic in spin can be neglected --- in particular in the
 linear approximation --- ECT is equivalent to GRT.
 To obtain essentially new effects, the density of spin squared
should   be comparable to the density of mass.
 For example, to achieve this, a
nucleon of mass \( m \) should be squeezed so that its radius \(
r_{\rm Cart} \) be such that
\begin{equation*}
\left( \frac{\ell^2}{r_{\rm
Cart}^{3}}\right)^{2}\approx \frac{m}{r_{\rm
Cart}^{3}}.
\end{equation*}
Introducing  the Compton wavelength \( r_{\rm Compt}=\ell^2/m\approx
10^{-13} \) cm, one can write
\begin{equation*}
r_{\rm Cart}\approx  (\ell^{2}r_{\rm Compt})^{1/3}.
\end{equation*}
The ``Cartan radius'' of the nucleon,
  \( r_{\rm Cart}\approx 10^{-26}\)
cm, so   small when compared to its physical radius under normal
conditions, is much larger than the  Planck length. Curiously
enough, the energy \( \ell^2/r_{\rm Cart} \) is of the order of the
energy at which, according to some estimates, the grand
unification of interactions is presumed to occur.

\smallskip

\noindent{\sffamily\bfseries Cosmology with spin and torsion}
\smallskip

In the presence of spinning matter, \( T^{\rm eff} \) need not
satisfy the
positive energy conditions,  even if \( T \) does.
 Therefore, the classical singularity  theorems of Penrose  and  Hawking
 can here be  overcome. In ECT,  there are simple
 cosmological solutions without  singularities. The simplest such solution,
 found in 1973 by Kopczy\'nski, is as follows.
 Consider a  Universe filled with a spinning dust
  such that \(P^\mu=\rho u^\mu\), \(u^{\mu}=\delta^{\mu}_{0} \),
 \( S_{23}=\sigma \), \( S_{\mu\nu}=0 \) for \( \mu+\nu\neq 5 \) and
 both \( \rho \) and \( \sigma \) are functions of \( t=x^{0} \)
alone. These assumptions are compatible with
 the
 Robertson--Walker line element \(\rmd t^{2}-\mathcal{R}(t)^{2}(\rmd x^{2}+ \rmd
y^{2}+
 \rmd z^{2})\), where \( (x,y,z)=(x^1,x^2,x^3) \) and torsion is determined
 from \eqref{e:Q=}. The Einstein equation \eqref{e:Ein} reduces to
the modified Friedmann equation,
 \begin{equation}\label{e:F}
\tfrac{1}{2} \dot{\mathcal{R}}^{2}-M\mathcal{R}^{-1}+
\tfrac{3}{2}S^{2}\mathcal{R}^{-4}=0,
 \end{equation}
supplemented by the conservation laws of mass and spin,
\begin{equation*}
M=\tfrac{4}{3}\pi\rho \mathcal{R}^{3}=\mbox{const.},\quad
S=\tfrac{4}{3}\pi\sigma \mathcal{R}^{3}=\mbox{const.}
\end{equation*}
 The last term on the left side of \eqref{e:F} plays the role of a
 repulsive potential, effective at small values of \( \mathcal{R} \); it
 prevents the solution from vanishing.
 It should be noted, however,
 that even a very  small  amount of shear in \( u \)  results in a
 term counteracting the repulsive potential due to spin.
 Neglecting shear and making  the (unrealistic) assumption that  matter in
 the Universe at \(t=0\)
 consists of about \(10^{80}\) nucleons of mass \(m\) with aligned spins, one
obtains the estimate \(\mathcal{R}(0)\approx 1\) cm and a density
of the order of \(m^2/\ell^4\), very large, but much smaller than
the Planck density \(1/\ell^2\).

Tafel (1975) found large classes of cosmological solutions with
a spinning fluid, admitting a  group of symmetries transitive on the
hypersurfaces of constant time. The models corresponding to
symmetries of Bianchi types I, VII\(_0\) and V are
non-singular, provided that the influence of spin exceeds that
of shear.
\newpage

\noindent{\sffamily\bfseries\large Summary}
\medskip

The Einstein-Cartan theory is a viable theory of
gravitation that differs very slightly from the Einstein theory;
the effects of spin and torsion can be significant only at
densities of matter that are very high, but nevertheless much
smaller than the Planck density at which quantum gravitational
effects are believed to dominate. It is possible that the
Einstein--Cartan theory will prove to be a better
classical  limit of a future quantum
theory of gravitation than the theory without torsion.
\medskip


\noindent {\sffamily\bfseries\large Further Reading}
 \medskip

\begin{harvard}

\item[]
Arkuszewski, W, Kopczy\'nski, W, and Ponomariev, VN (1974).
\newblock On the linearized Einstein--Cartan theory.
\newblock {\em Ann. Inst. Henri Poincar\'e\/} 21: 89--95.

\item[]
Bailey, I and Israel, W (1975).
\newblock Lagrangian dynamics of spinning particles and
polarized media in general relativity.
\newblock {\em Commun. Math. Phys.\/} 42: 65--82.

\item[]
 Cartan, \'E (1923), (1924) and (1925).
\newblock Sur les vari\'et\'es \`a connexion affine et la th\'eorie de
la relativit\'e g\'en\'eralis\'ee.
\newblock Part I: {\em Ann. \'Ec. Norm.\/} 40: 325--412 and ibid. 41: 1--25;
Part II: ibid. 42: 17--88; English transl. by A Magnon and A Ashtekar,
{\em On manifolds with an affine connection and the theory of general
relativity}.
\newblock Napoli: Bibliopolis (1986).

\item[]
Cosserat, E and F (1909).
\newblock {\em Th\'eorie des corps d\'eformables\/}.
\newblock Paris: Hermann.

\item[]
Hammond, RT (2002).
\newblock Torsion gravity.
\newblock {\em Rep. Prog. Phys.\/} 65: 599--649.

\item[]
Hehl, FW, von der Heyde, P, Kerlick, GD and Nester, JM (1976).
\newblock General relativity with spin and torsion: Foundations
and prospects.
\newblock {\em Rev. Mod. Phys.\/} 48: 393--416.

\item[]
Hehl, FW, McCrea, JD, Mielke, EW and Ne'eman, Y (1995).
\newblock Metric-affine gauge theory of gravity: field equations,
Noether identities, world spinors, and breaking of
dilation invariance.
\newblock {\em Phys. Reports\/} 258: 1--171.

\item[]
 Kibble, TWB (1961).
\newblock Lorentz invariance and the gravitational field.
\newblock {\em J. Math. Phys.\/} 2: 212--221.

\item[]
Kopczy\'nski, W (1975).
The Palatini principle with constraints.
\newblock {\em Bull. Acad. Polon. Sci., s\'er. sci. math. astr.
 phys.\/} 23: 467--473.

\item[] Mathisson, M (1937).
\newblock Neue Mechanik materieller Systeme.
\newblock {\em Acta Phys. Polon.\/} 6: 163--200.

\item[]
Penrose, R (1983).
\newblock Spinors and torsion in general relativity.
\newblock {\em Found. of Phys.\/} 13: 325--339.

\item[]
Ruggiero, ML and Tartaglia, A (2003).
\newblock Einstein--Cartan theory as a theory of defects in space-time.
\newblock {\em Amer. J. Phys.\/} 71: 1303--1313.

\item[]
 Sciama, DW (1962).
\newblock On the analogy between charge and spin in general relativity.
\newblock In: (volume dedicated to L Infeld)
{\em Recent Developments in General Relativity\/}, pp 415--439.
\newblock Oxford: Pergamon Press  and Warszawa: PWN.

\item[] Tafel, J (1975).
\newblock A class of cosmological models with torsion and spin.
\newblock {\em Acta Phys. Polon.\/} B6: 537--554.

\item[]
Trautman, A (1973).
\newblock On the structure of the Einstein--Cartan equations.
\newblock {\em Symp. Math.\/} 12: 139--162.

\item[] Van Nieuwenhuizen,  P (1981)  Supergravity.  {\em Physics Reports\/}
 68: 189--398.
\end{harvard}
\end{document}